# A Framework for institutional change in the age of AI

David Perl-Nussbaum & Noah D. Finkelstein
Department of Physics, University of Colorado Boulder

**Abstract**

Generative AI is rapidly reshaping STEM higher education. Not only are our educational practices changing, but how we think about educational transformation must adapt. Existing models of institutional change in STEM, aimed at interactive engagement, have largely followed an *adoption* logic: relatively stable, well-researched educational practices are evaluated and then scaled. These assumptions do not hold for generative AI, which is an *arrival technology* – entering classrooms before a sufficient pedagogical evidence base could form. Building on recent decades of work on STEM education institutional change, we propose a framework identifying six dimensions along which prior change models must be reconsidered in light of AI: three concerning the tools at the center of reform (the tool's evidence base, rate of change, and scope), and three concerning the people involved in change (faculty, change agents, and students). For each dimension, we examine how AI-era assumptions differ from those underlying prior STEM education reforms and derive design implications, including: privileging humble and local inquiries; organizing reform around pedagogical approaches rather than specific tools; repositioning change agents as facilitators of collective inquiry; and engaging students as partners in reform. Collectively, the six dimensions and design implications constitute a new framework for adapting change models to support institutions under conditions of genuine uncertainty. Finally, we illustrate how the framework may be applied through a brief case-study of a faculty workshop series carried out in a university physics department to support instructors adapting to this modern AI era.

**keywords:** *Institutional change; generative AI; change theory; arrival technology; interactive engagement; STEM higher education; professional development; change agents.*

## 1. Introduction

The arrival of generative artificial intelligence (AI) is transforming our routines of learning, teaching, and assessing, creating an urgent need to lead thoughtful institutional responses to AI (Hughes et al., 2025; Walczak & Cellary, 2023). As educators, our goals remain what they have been: to promote student-centered, evidence-based approaches to learning in STEM fields. To pursue these goals at scale, the field of institutional change has produced valuable models for transforming departments and institutions directed at the use of these evidence-based, interactive engagement (IE) and active learning techniques (ASCN, 2021; NASEM, 2025; White et al., 2021).

However, simply applying models developed to promote IE practices is now insufficient in an AI-infused world. IE reforms[1] share a critical feature: they were built on an established evidence base that justified specific innovations before scaling them. Educational innovations such as *Peer Instruction* (Mazur, 1997), *Tutorials in Introductory Physics* (McDermott & Shaffer, 2002), the *Learning Assistant* program (Otero et al., 2006), and interactive simulations (Wieman et al., 2008) were designed with learning goals in mind, evaluated, and made available to faculty alongside evidence of their effectiveness (Chasteen et al., 2015; Hake, 1998; Wieman, 2017). Reich and Dukes (2025) label technologies that follow this pattern *adoption technologies*.

AI-based practices do not fit this pattern. They entered our classrooms not through slow and intended adoption of evaluated innovations, but through sudden and widespread use – largely before institutions had formed any response (Alba et al., 2026; Xiao et al., 2023). Reich and Dukes (2025) call such technologies *arrival technologies*: they disrupt our practices before we understand them. This is not the first time outside created tools have arrived to impact our classrooms. While prior technologies such as mobile phones or calculators share some arrival technology features, they appeared slowly and at longer cycles of iteration – in the case of the calculator, at roughly ten-year intervals (the handheld calculator, the scientific calculator, the graphing calculator, web-based platforms) – which allowed educators and schools to adapt. Generative AI offers no such interval. Reich & Dukes (2025) argue that ChatGPT in this sense is unique - the first widely and instantly embedded *arrival technology*. It may be the first in “a new cadre of arrival technologies” (Reich & Dukes, 2025, p.22), challenging educators and institutions as they leave no interval for adaptation.

Our IE goals to support student learning and engagement may not have changed; but the conditions and mechanisms certainly have. Moreover, because students will enter professions increasingly shaped by AI, learning to engage with it critically and productively is itself now an IE concern – an extension of our longstanding commitment to student agency and authentic disciplinary practice.

[1] Here we do not distinguish between reforms directed at promoting interactive engagement (IE), active learning, inclusive pedagogy, use of research-based instructional strategies (RBIS), evidence-based instructional practices (EBIP), etc. We will use IE as a proxy for the breadth of such efforts. While they do differ and have different historical roots and communities, they hold in common the attributes across the framework introduced below.

The pedagogical evidence base that justified IE reforms does not yet exist for the AI age. Despite the proliferation of AI literacy frameworks, it may be years until we understand what productive, pedagogically sound use of AI in disciplinary learning looks like. We cannot wait for best practices to emerge before transforming our teaching, as our courses have already been transformed by students' use, campus policies, and individual faculty deployment of AI. This creates the central dilemma that this paper addresses: institutions cannot wait to formulate AI responses – as students are already heavily relying on AI tools, but neither can they responsibly scale innovations that remain unjustified. Neither banning nor uncritically embracing generative AI is tenable (Alba et al., 2026; Guest et al., 2025; Xiao et al., 2023). **What is needed is a model for leading institutional change under conditions of genuine uncertainty.**

As institutions wrestle with their response to generative AI, the question of how to lead that response remains undertheorized. Missing from both the institutional change literature and the AI-in-education literature are frameworks for institutions to initiate, coordinate, and sustain change when the pedagogical evidence base is still forming. We address this gap by offering a framework that acts with *intellectual humility* (Pritchard, 2020) – aiming for thoughtful change without overclaiming what we know (Reich & Dukes, 2025).

Our framework identifies six dimensions along which existing IE change models must be adapted to remain effective in the AI era. Three dimensions concern the tools at the center of change – their evidence base, rate of change, and scope of intended use. Three dimensions concern the human actors – faculty agency, change agent role, and student role. For each dimension, the framework examines the state of existing IE change models, and the AI-specific characteristics that demand reconsideration. In some dimensions, existing strategies translate directly; in others, their assumptions must be substantially rethought.

This paper makes three contributions. First, we identify and examine six dimensions along which AI-driven institutional change must be reconsidered relative to prior IE reforms. Second, we derive design implications from these dimensions to guide AI-driven change initiatives. Third, we demonstrate the framework's practical utility through a case study of a faculty workshop series conducted in our physics department at the University of Colorado Boulder.

## 2. Institutional Change in STEM Higher Education

Research on institutional change in higher education has produced a rich body of theory alongside practical models and strategies for change. We briefly review dominant theoretical frameworks and models that have guided past STEM education reforms. Rather than seeking to be exhaustive (see Reinholz et al., 2021, for a systematic review), we identify core assumptions embedded in these frameworks – assumptions we argue should be reconsidered when planning AI-driven change.

### 2.1 Change theories

Researchers have turned to theories from various fields to inform their design, implementation and research of educational change efforts (Reinholz et al., 2021). As a main aim of these efforts is the adoption and scaling of evidence-based instructional practices, a common starting point is Rogers' (2003) *Diffusion of Innovations* (DoI) theory.

DoI describes how innovations, whether new ideas, practices, or technologies, spread through a social system, moving through five stages – knowledge, persuasion, decision, implementation, and confirmation – and across adopters ranging from innovators and early adopters to laggards. The theory identifies innovation characteristics that predict successful diffusion: relative advantage over existing practices, compatibility with existing practices, low complexity, ease of trial (trialability), and observability of outcomes. Diffusion is further shaped by *communication channels* and through *change agents* – individuals who actively promote adoption within the social system.

DoI is appealing in the context of STEM education reform because it offers a vocabulary for thinking about why some innovations spread and others do not, and what shapes the pathways and pace of diffusion. However, the model rests on the assumption that the innovation in question is sufficiently stable and well-defined to be evaluated, adopted, and diffused. As Rogers (2003) notes, the innovation-diffusion process is fundamentally an "uncertainty reduction process" (p. 232): it works when uncertainty can be reduced through accumulated evidence. This assumption, we argue, does not hold for generative AI, which entered educational systems before it could be stabilized or evaluated.

Indeed, Reinholz and colleagues (2021) found that DoI is a commonly used theory underlying many educational change efforts. Evidence, however, clearly shows that approaches that draw from such linear models of change, e.g., dissemination of materials or "top-down" policies, fail to achieve sustainable change (Dancy & Henderson, 2008; Henderson et al., 2011; Reinholz & Apkarian, 2018). This has led to the understanding that we should move beyond models focused on individual adopters to organizational theories that address the complexity of academic institutions, including social, cultural and political perspectives (Kezar, 2018).

A prominent model adopted from organizational change literature to the context of higher education change is Bolman and Deal's (2017) four-frame model, which provides four interrelated lenses for understanding organizations: structural (policies, rules, arrangements), human resources (identities, needs, experiences of people), political (formal power and informal influence), and symbolic (culture, meaning, values). Applied to higher education change efforts, this framing helps change leaders make sense of a complex system and identify which levers are available in their specific context as well as help them attend to multiple frames simultaneously and resist the tendency to rely on a single frame (Laursen & Austin, 2020; Reinholz & Apkarian, 2018).

In her book, Kezar (2018) offers a wider lens on institutional change. She synthesizes organizational change literature to provide six change perspectives that differ in their conceptualization of change and in their view on how and why change occurs. These include *scientific management*, *evolutionary*, *political*, *social cognition*, *cultural*, and *institutional*. For example, through the *evolutionary* perspective, change is unplanned, resulting from complex systems adapting to external pressures and changing environments. Under this perspective, the role of leaders is to manage and respond to inevitable changes, and create structures that are flexible and thus responsive to external forces (see Corbo et al., 2015). As another example, the *social cognition* perspective holds that change can be achieved only when individuals shift their thought processes. Accordingly, change agents should provide opportunities for sensemaking about the desired change, e.g. helping faculty explicate their underlying beliefs and confronting them with institutionally collected data. According to Kezar (2018) and Bolman and Deal (2017), change agents typically adopt a single frame or perspective to induce change, often missing the complexity of the system.

Systemic approaches offer a more adequate account of institutional complexity than linear diffusion models. Yet, they largely share a common feature relevant to our argument: they are models for navigating change toward a known destination (i.e., interactive engagement). The dominant approach used in IE reform (with the notable exception of Kezar's evolutionary perspective) presuppose that the direction of change is justified in advance and the innovations – whether teaching practices or educational tools – are evidence-based. It is this presupposition that the arrival of generative AI fundamentally disrupts.

### 2.2 Change initiatives

Ultimately, we seek to put theory into practice to impact the lives and experiences of learners in STEM higher education. In a seminal review paper, Henderson and colleagues (2011) analyzed 191 STEM education transformation articles. They categorized their change approaches along two axes: whether the change targets individuals or systems, and whether desired outcomes are prescribed or emergent. This resulted in four categories of change approaches: disseminating curriculum and pedagogy (individual / prescribed); developing reflective teachers (individual / emergent); enacting policy (environment / prescriptive); and developing shared vision (environment / emergent). The analysis also pointed to effective and sustainable change strategies, including long-term efforts, focus on changing faculty conceptions, and taking departmental and institutional culture into account.

Henderson et al. (2011) highlight the need to work across all four quadrants. Indeed, the field of faculty development seeks to develop capacities for navigating emergent phenomena, and reforms aimed at developing shared vision and cultural change can also be emergent. However, STEM education has been dominantly working at the prescriptive side of change. This reflects the field's strength: discipline-based education research produces evaluable instructional innovations, such as *Peer Instruction* or *Tutorials*, and the logic of reform follows accordingly, explicitly or implicitly drawing from Rogers' (2003) DoI model (Reinholz et al., 2021).

Subsequent change initiatives have advanced along different axes. The Science Education Initiative (SEI), which is widely considered a successful model for reform, shifted from targeting individuals to targeting environments (Chasteen et al., 2015; 2016). It funded departments to employ disciplinary Science Teaching Fellows (STFs) – discipline-based postdocs focused on education who partnered with faculty on course transformation, developing learning goals, identifying student difficulties, and creating materials and assessments aligned with evidence-based practices. The SEI enhanced partnerships among faculty and other departmental change agents, attending to departmental norms, relationships, and structures.

Building on this SEI model, the Departmental Action Team (DAT, Corbo et al., 2015; Reinholz et al., 2019) and Departmental and Leadership Teams for Action (DeLTA , Andrews et al., 2021) approaches moved toward the emergent end of change, organizing faculty into working groups that collectively identify problems and pursue change through shared vision and iterative inquiry. These models represent genuine advances. Yet, even the most emergent among them operates within a landscape where the broader objective of change – active learning and evidence-based pedagogy – is established in advance. These emergent models remain highly relevant for AI-driven change, but arrival technologies and AI tools intensify particular dimensions of uncertainty because the technology is externally introduced, rapidly changing, general-purpose, student-led, and not yet pedagogically stabilized. Some elements from past IE change models remain useful, while others should be rethought and developed.

In the following section, we develop our framework for leading institutional change in the age of AI. We draw from the change initiatives and underpinning theories reviewed here to consider what remains the same and, more importantly, what is fundamentally different for AI-based transformations.

## 3. The Framework: Six Dimensions for Adapting IE Change Models to the AI Era

Our framework identifies six dimensions along which existing IE change models must be reconsidered in light of generative AI's arrival. The framework is motivated by sociocultural models of human activity systems (Cole 1996; Engeström, 1999), which define the central features of human activity as comprised of *people* engaged in an action directed at a given *object*, mediated by *tools*. Tools are not neutral instruments -- they transform the activity itself. The *object* is the aim of the activity. In STEM reform, this includes a focus on students' conceptual understanding, reasoning, and engagement. We offer our framework as a lens to consider how to pursue these same educational objectives in the AI era, where the *tools* and *people*'s engagement have drastically changed and transformed our activity systems. While the object of educational practices may change (necessarily or by choice) with the use of AI, here we separate the *object* of change (our goals of educational practice), from the roles of *people* engaged and the *tools* we use to accomplish the stated objectives.

Figure 1 provides an overview of the framework. The first three dimensions of our framework concern the *tools* - educational innovations at the center of change: their evidence base, rate of change, and scope. The other three concern *people* - the actors involved in the change process: faculty agency, the role of change agents, and the role of students. For each dimension we describe, we suggest *design implications* - what AI characteristics imply for the development of effective change initiatives.

**Six Dimensions for Rethinking Institutional Change Models in Response to AI**

| Traditional IE change | | AI-influenced change |
| --- | --- | --- |
| | **The Tools** | |
| *R&D before scaling*<br>Evidence reduces uncertainty. | Evidence Base | *Scaling before R&D*<br>Acting amid uncertainty. |
| *Stable tools*<br>Aiming for sustainability and transferability of materials. | Rate of Change | *Tools rapidly changing*<br>May be obsolete before dissemination (or within a semester). |
| *Designed for education*<br>Designed by and for educators. | Scope and Intended Use | *General-purpose tools*<br>Tools built by others for other uses. Coming to education from outside. |
| | **The People** | |
| *Faculty opt in*<br>Faculty are agents of choosing to use (or not use) IE practices. | Faculty Agency | *Forced upon faculty*<br>Either by institution or because of student use. |
| *Broker of knowledge and practices*<br>Additional person to drive, foster and support change. | Change Agent Role | *Lack of AI educational experts*<br>Status quo already disrupted. Must respond after the fact, without established practice to broker. |
| *Student use guided by faculty*<br>Students as endpoint of reform -- to be convinced of utility of change for their learning. | Student Role | *Students lead in use*<br>Students already engaged in use and seeking guidance. |

*Figure 1: Six dimensions – three concerning tools and three concerning people – along which institutional change in the AI era diverges from prior IE reforms.*

Key considerations in any educational reform practice are matters of ethics and equity (Reinholz et al., 2021), concerns that generative AI further compounds (Varsik & Vosberg, 2024). This framework supports attention to these essential matters across each of the six dimensions. While one could have a framework that calls out matters of ethics and equity as separate components, we opt to include these as underpinning principles and commitments that change agents may apply across the following dimensions.

## THE TOOLS

### 3.1 Dimension 1: Evidence base

This dimension concerns what is known about the educational innovation at the point when change is initiated. It is the foundational dimension in our framework, underlying all others, because the status of the evidence base shapes what kind of change process is warranted.

IE reform efforts begin with evidence-based and validated tools for widespread adoption (Hake, 1998; Wieman, 2017), thus following the adoption technology pattern (Reich & Dukes, 2025): research and development precede scaling. The sudden arrival of Generative AI inverts this sequence:

> *"ChatGPT arrived in schools without evaluation, assessment of risks and benefits, training for educators, or any steps considered essential to effective technology integration."* (Reich & Dukes, 2025, p.20)

AI enters our classrooms not through deliberate adoption of evidence-based tools but through widespread and spontaneous student and teacher use (heavily promoted by tech-companies), largely before institutions had formed any response (Hughes et al., 2025; Malik et al., 2023; Alba et al., 2026), and definitely before research is able to figure out productive and responsible practices for learning with AI (Klopfer et al., 2024). In some cases, the institutions themselves actively promote wide integration of unevidenced AI tools and practices (Jin et al., 2025).

The direct evidence base for effective use of AI-based pedagogy that could guide the use of such tools does not yet exist in recognizable form (Giannakos et al., 2025; Stracke, 2024) and may not for years. The challenge is not simply that we lack evidence – it is that we as educators are called to act, either by external pressures or because of widespread student use disrupting our established instructional routines before the foundational questions have been addressed. And in cases where we might find evidence for effective use of a given tool, then the tool itself changes (see next section). This creates a genuine dilemma. Some, drawing on adoption frameworks, proceed as though best practices for AI-enhanced learning are established and ready for dissemination. Others argue for waiting until the evidence is in (Xiao et al., 2023). We argue that neither position is tenable. The first overclaims what we know; the second ignores that student learning is already being reshaped by AI, with or without deliberative involvement. We argue that applying existing change frameworks to generative AI is misaligned: adoption models presuppose the kind of evidence base that arrival technologies, and AI especially, lack.

**Design implications:**

*Humble inquiries:* Here we join Reich & Dukes (2025) in calling for humble inquiries and acting with epistemic, or *intellectual humility* (Pritchard, 2020). Rather than advocating or scaling practices whose effectiveness has not been established, we should seek out and learn from localized uses of AI that show promise – surfacing what faculty are already trying, documenting what appears to work and for whom, and sharing these emerging cases without overclaiming their generalizability. Hopefully, these collective efforts will aggregate to later establish the evidence for developing future research-based instructional practices for productive use of AI in learning. While it may be useful to draw from general AI literacy frameworks (Anthropic, 2026; Hibbert et al., 2024; OECD, 2025) that give us some purchase into the use of these tools, we ought to do so with caution and tentatively, as they do not yet rest on well-established data (Giannakos et al., 2025). Scaling unevidenced practices is also an equity concern, risking exacerbation of existing gaps rather than narrowing them (Jamal Eddine et al., 2026), as learners with greater resources and digital literacy are best positioned to benefit from early-stage efforts.

In terms of institutions, this means providing resources for experimentation cycles and gaining familiarity with AI, and providing access to AI (with guardrails), without making long-lasting pedagogical or financial commitments (Klopfer et al., 2024). The approach of distributing studies and approaches across an institution, along with adequate support and recognition to do so, and mechanisms for sharing outcomes and discerning what is generalizable holds some promise. Such approaches will support those individuals engaging in the practices and build collective wisdom across the community that is relevant to the local environments (departments and institutions).

This form of local inquiry is not new. IE reforms, particularly the SEI and related models, embedded systematic investigation into the change process itself – faculty and change agents identified student difficulties, revised learning goals and courses based on locally collected evidence (Chasteen et al., 2015). We argue that this same practice of grounded, course-level inquiry is essential for AI-driven change, though its purpose shifts. Where IE inquiries helped faculty select and implement evidence-based practices, AI inquiries are exploratory, aimed at generating insights. Furthermore, AI change efforts should draw on IE and learning theory to identify areas to focus on – including student engagement, authentic practice – and explore how AI might be integrated into or alongside these proven approaches.

### 3.2 Dimension 2: Rate of change

Where Dimension 1 asks what is known about an innovation, this dimension asks how long that knowledge will remain relevant. The two are related but distinct: an innovation could have a weak evidence base yet be stable enough for evidence to accumulate, or be well-evidenced initially but evolve so quickly that the evidence becomes obsolete. Generative AI suffers from both constraints: inadequate foundational evidence compounded by change too rapid for knowledge to solidify.

IE reforms typically operate under conditions of tool stability. Once an educational innovation such as clickers or Tutorials is introduced, the tool itself is not expected to change substantially over the course of a reform effort. This stability is what makes adoption and scaling coherent – the object being scaled is the same as the object that was studied. Sustainability and transferability of instructional materials across instructors and institutions thus become explicit measures of success, and curating stable materials is a valuable investment (Chasteen & Perkins, 2014).

AI tools follow a different pattern. Contrary to early predictions of a plateau, the 2026 AI Index report finds that AI capabilities are still accelerating (AI Index Steering Committee, Sajadieh et al., 2026), with frontier models gaining new capabilities on timescales of months rather than years. AI also shows what is often called a "jagged frontier" (Dell'Acqua et al., 2023): AI models handle some tasks well and fail at others that seem similar, and the line between the two moves with each release.

Conventional research cycles are already struggling to keep pace with practitioner needs, and the gap is sharper with AI: findings about specific tools risk obsolescence before reaching publication (Baytas & Ruediger, 2025; Holmes & Tuomi, 2022; Hughes et al., 2025; Kortemeyer et al., 2025). Faculty encounter this instability even more acutely: within a single semester, a tool introduced in Week 1 may behave meaningfully differently by Week 14.

**Design implications:**

*Approaches over tools.* AI driven change should focus on *approaches* rather than *tools* – pedagogical orientations and practices that can survive the replacement of any particular model (e.g., validation and refinement, Ben-Zion et al., 2026). Where SEI and related models of change could reasonably aim to build a repository of course materials, AI-driven change should aim to build a repository of approaches and related example cases. These approaches should draw on the stable evidence base from IE reform – e.g., what we know about student engagement, authentic practice, and epistemology in STEM education does not become obsolete when a new model is released. This reframing also shifts what institutions should invest in – from tool procurement (e.g. particular grading systems or tutoring agents) toward sustained support for faculty inquiry and iteration.

*Evolving guidelines.* Because change occurs even within semesters, course policies around AI use should not be handed down as fixed rules but developed through ongoing dialogue between instructors and students. And setting classroom culture where students participate in articulating what responsible use looks like for a given course and classroom. Policies should be revisited and iterated as tools evolve and as students and faculty develop shared experience (see also Dimension 6). Yet as Klopfer et al. (2024) caution – an iterative stance cannot excuse us from coherent policy: students and faculty need recognizable principles around ethics and academic integrity, even as specifics evolve.

### 3.3 Dimension 3: Scope and intended use

Finally with respect to the tools (AI or IE approaches), we must consider the purpose for which the tools for educational reform were developed – whether purpose-built for education or developed as a “general purpose” technology and then appropriated for educational use. This distinction determines who has interest in promoting adoption, the external pressure institutions face to use it, and the ethical questions that accompany adoption in the classroom.

IE innovations were, by design, educational tools. Clickers, Tutorials, Peer Instruction, and interactive simulations were developed by educators and education researchers with specific learning goals and education theory embedded in their design, (Mazur, 1999; McDermott & Shaffer, 2002; Otero et al., 2006; Wieman et al., 2008). Accordingly, debate around their adoption was largely pedagogical and internal to the educational community: e.g., focusing on questions about implementation or fidelity (Turpen & Finkelstein, 2009).

Generative AI was not designed for education; it was developed as a general-purpose technology (Bresnahan & Trajtenberg, 1995; Calvino et al., 2025), and has entered educational contexts both through widespread individual use and active vendor promotion. Some of its common uses – producing finished work with minimal effort – can stand in direct tension with the purposes of learning (Giannakos et al., 2025; Messeri & Crockett, 2024; Wulff & Kubsch, 2025). Because AI originates outside education and operates at societal scale, its adoption is inseparable from broader public debate. AI tools arrive with hype, ethical controversy, concerns about equity and environment, and the commercial interests of the companies producing them (Bond et al., 2024). Students encounter these tools across every domain of their lives, not only in coursework, and many uses have already become normalized and effectively invisible – embedded in search engines and writing software – without users recognizing an AI layer is present (Xiao et al., 2025).

Unlike traditional IE tools, which functioned purely as mediators through which students engaged with disciplinary content, AI is simultaneously a mediator and an object of learning: students must learn to use it productively, and the disciplinary practices we teach are themselves being reshaped by it (Wang et al., 2023). Faculty are therefore not simply adopting a new tool in service of existing learning goals, but renegotiating what those goals are (Engeström, 1999; Finkelstein, 2025).

**Design implications:**

*Broadening Goals.* Faculty engaging in AI-driven change need to be prepared – and institutionally supported – to educate students beyond disciplinary content to address ethics, agency, and the role of AI in students' disciplinary futures (Baytas & Ruediger, 2025). These are not distractions from disciplinary teaching; given that AI is simultaneously a mediator and an object of learning, they are inseparable from the work of teaching the discipline itself. This represents a genuine expansion of what faculty are being asked to do and should be recognized as such. Faculty need to partner with institution leaders (and in so far as possible with those constructing technologies for education) to inform where and how such tools are applied to our environments.

*Flexible and general tools over dedicated platforms.* Dedicated platforms carry obsolescence risk (see Dimension 2). A more durable strategy, and one which is more aligned with students' broader societal needs, would be to help students develop effective practices for using these general tools critically – practices that transfer across contexts from STEM learning to everyday life (Hibbert et al., 2024; OECD, 2025). Rather than procuring external solutions, institutions should invest in supporting faculty and students to collaboratively adapt general tools to their local educational contexts, building capacity from within. This is also, arguably, the more natural direction: students are likely to reach for general tools already embedded in their lives rather than navigating multiple course-specific platforms (Reich & Dukes, 2025).

## THE PEOPLE

### 3.4 Dimension 4: Faculty agency

Another dimension that varies in change strategies is the degree to which faculty choose to engage with an educational change effort – and what that choice, or its absence, means for how institutional change should be designed.

In most IE reforms, faculty opt-in. For example, in the SEI initiative, a central team recruited faculty who had expressed interest in transforming their courses, and partnered with them to transform their courses (Chasteen & Perkins, 2014; Chasteen et al., 2015). Faculty participation was voluntary, and the motivation driving it was largely intrinsic, including interest in student learning and dissatisfaction with their current teaching methods (Andrews & Lemons, 2015; Biswas et al., 2022). Faculty can decide if and how they want to transform their courses (Dancy et al., 2016) and whether to adopt supporting structures like the Learning Assistant model (Otero et al., 2006) – with participation remaining voluntary throughout. Of course, for sustained engagement, institutions and change agents need to provide external and systemic resources.

For many faculty, the decision to engage with generative AI has been made for them – whether by institutional policies, or from the widespread use of students (Digital Education Council, 2024). In a sense, their course has already been transformed, whether faculty are aware of it or not. The question is no longer whether to engage in course transformation, but how to respond to the changes that already

occurred – students are using AI tools, largely in an unsupervised fashion. This creates deep concerns among faculty (Hughes et al., 2025; Watson & Rainie, 2025), who feel unequipped to deal with this sudden disruption and did not have voice in when and how these tools arrived. This concern may also be an opportunity where faculty are more open to change and receptive to support if they are "dissatisfied" with their current course methods/outcomes (Andrews & Lemons, 2015).

The uninvited appearance of AI in our classes complicates faculty motivations. Many faculty members are encountering disruption they did not choose and were not prepared for. At the same time, institutional pressures are creating a different form of forced engagement: some departments and institutions are mandating responses to AI, or procuring AI tools for instructional use – sometimes without genuine faculty input.

Importantly, the degree of disruption also varies considerably across faculty and instructional contexts. Upper-level and graduate courses may be less affected by AI than large introductory courses where AI use is widespread and difficult to detect. Writing and computational courses may be heavily disrupted, while laboratory and experimental courses may be less disrupted. A change framework that treats all faculty as equally affected misses this heterogeneity and risks irrelevance or unnecessary burden for those who experience little disruption, while under-resourcing those who face the most significant challenges.

**Design implications:**

*Map disruption and needs.* AI-driven change meets faculty at widely different starting points – from active experimentation to anxiety or despair. Effective support must therefore begin by surfacing where faculty actually are: their concerns, sense of disruption, and the degree to which their course structures and teaching practices have been impacted by AI arrival. Structured faculty interviews or departmental conversations can inform differentiated support structures.

*Invest in support structures.* Because many faculty are engaging with AI under conditions they did not choose, institutional structures should provide spaces for collective inquiry and resources for response. Here the SEI offers a direct precedent: rather than expecting faculty to reform their courses in addition to their other obligations, it funded discipline-based Science Teaching Fellows to serve as genuine partners in course transformation (Chasteen et al., 2016). An analogous investment is warranted for AI-driven change: not workshops on available AI tools, but sustained spaces for collective faculty inquiry and dedicated experts who can help faculty surface what is happening in their courses and think through responses. As we have experienced - collective, facilitated discussions can be a valuable opportunity to rethink not only our practices, but also our disciplinary learning goals which may be disrupted by these new technologies.

### 3.5 Dimension 5: The role of change agents

While embedded experts can serve as facilitators of institutional reform (Huber & Hutchings, 2021), expertise in AI-based educational practices is largely absent; our facilitators need to serve different roles in AI induced educational transformation.

In SEI and related models, embedded experts occupy a bridging role, linking between the knowledge produced by discipline-based education research and the faculty interested in course transformation, and then worked alongside faculty to assess these transformations through local evidence (Andrews & Lemons, 2015; Biswas et al., 2022; Chasteen et al., 2016).

With generative AI, change agents cannot serve as brokers of known tools or best practices to faculty. These individuals supporting AI reform cannot reduce uncertainty in the way Rogers (2003) envisions, because the uncertainty in practice itself is not yet reducible. The role of change agents must therefore be rethought.

In parallel, the forms of expertise of these change agents need to be reconsidered. In the prior reforms, embedded experts were selected primarily for disciplinary knowledge, educational awareness and interest, and interpersonal skills. DBER-trained postdocs were scarce, but also less necessary given that research-based practices were already developed and agreed upon (Chasteen et al., 2015). The relevant expertise was in facilitating adoption and managing relationships, not in producing the pedagogical knowledge itself. For AI-driven change, the question becomes sharper: should STFs be disciplinary experts, AI experts, or experts in pedagogy and learning? We caution against overly valuing AI tool expertise. For example, change agents drawn from industry or vendor contexts may be implicitly oriented toward promoting adoption rather than critically evaluating it. Pedagogical and disciplinary expertise – grounded in what we know about student learning, epistemology, and authentic practice in STEM – is arguably more durable and would allow such change agents to engage with faculty in local inquiries and experimentation with new tools and practices.

**Design implications:**

*Change agents as facilitators of collective inquiry.* We argue that we should follow an embedded expert model (Huber & Hutchings, 2021) and hire "AI STFs"; however, rather than brokering, AI-oriented change agents should identify where faculty and courses are most disrupted and facilitate collaborative inquiry and design of thoughtful responses. One possible venue for gathering local data and developing a shared vision (Henderson et al., 2011), is through facilitating faculty Departmental Action Teams or professional learning communities (Reinholz et al., 2019). Because neither the change agents nor the faculty possess the relevant expertise yet, the work is inherently collective.

*Prioritize pedagogical expertise.* Institutions seeking to hire or train embedded change agents should resist the pull toward AI-expertise as a primary qualification. Deep familiarity with learning theory, disciplinary practices, and DBER positions change agents to evaluate AI uses against trusted pedagogical standards. Familiarity with AI tools is essential and important, and can be developed over time; it should not substitute for the kind of educational grounding that supports our foundational goals of developing students.

### 3.6 Dimension 6: Student Role

The final essential human dimension to consider in light of the modern AI landscape is the role of students in the change efforts. Students are essential actors who actively shape the goals and direction of change.

Often in IE reforms, students are considered the endpoint of the change effort, but rarely participants in it. For example, in the Henderson et al. (2011) framework - change efforts are organized along axes of systems vs individuals – meaning faculty. Students motivate reform; but they do not participate in the change (other than perhaps serve as research subjects to inform the object of change). There are some educational transformation initiatives that actively engage students in the change practices (e.g., the LA model, Otero et al., 2006; the PLTL model, Wilson & Varma-Nelson, 2016). In such models, our students serve not only as a mechanism for enacting reforms, but also shape what these practices are, through partnership with faculty who are engaging in student-centered educational practices (Abdurrahman et al., 2022).

With generative AI, students lead in use. Survey data consistently reveals students' widespread use of AI for learning, and that students actively seek guidance on responsible, productive use of AI (Digital Education Council, 2024; Watson & Rainie, 2025). Students are not awaiting reformed instruction, but are already experimenting, and their practices and goals should be considered crucial inputs for change efforts. In many senses, the students are the initiators of the AI-spurred educational changes that are happening; however, they are largely doing so in an unsupervised fashion and have differential access and preparation for AI use. There is also a deep concern that is noted by our students and instructors alike: In an era when AI systems can increasingly perform cognitive tasks that once required human effort, deliberately maintaining students as active and reflective agents in the construction of their own learning may be among our most important learning goals (Author, 2025). Finally, the ethical concerns around environmental and social impact have largely gone unaddressed.  Including student (and faculty voice) to these ends will be necessary.

**Design implications:**

*Students as partners.* AI-driven institutional change should position students not simply as the recipients of reform but as genuine participants in shaping it (Cook-Sather & Matthews, 2021). This means understanding students' current AI use, and involving them in conversations about which learning goals are worth protecting, and what productive AI practice should look like in their disciplines. Students have a particular stake in these questions as they are the ones entering professions increasingly shaped by AI. Because neither faculty nor institutions yet possess settled answers to these questions, students' experience of the evolving landscape is an essential input to the change process (Bond et al., 2024; Ripley et al., 2024).

*Build on, and redirect, student use.* Rather than acting as though students are blank slates and AI is not in use already, change efforts should take seriously the beliefs and practices students bring to our classes. This requires surfacing those practices, understanding the goals motivating them, and designing course experiences and institutional supports that engage with those goals -- redirecting student energy toward productive and critical use rather than prohibiting or ignoring existing patterns. We found a structured survey of students' current AI use, administered early in a course, to be a useful entry point: the data informs instructors' understanding of student practice, and can serve as the basis for collaborative discussions around AI norms and policies. We return to this in Section 4.

*Equitable and ethical use of AI:* As AI becomes prevalent across higher educational institutions, we must redouble our efforts to ensure universal access to these essential tools, while providing safeguards for privacy and personal protection, and addressing structural biases. Finally, institutions leaders need to partner with students in making choices about the ecological and social implications of their explicit adoption of these tools.

Figure 2 provides a summary of the framework, including the six dimensions and derived design implications.

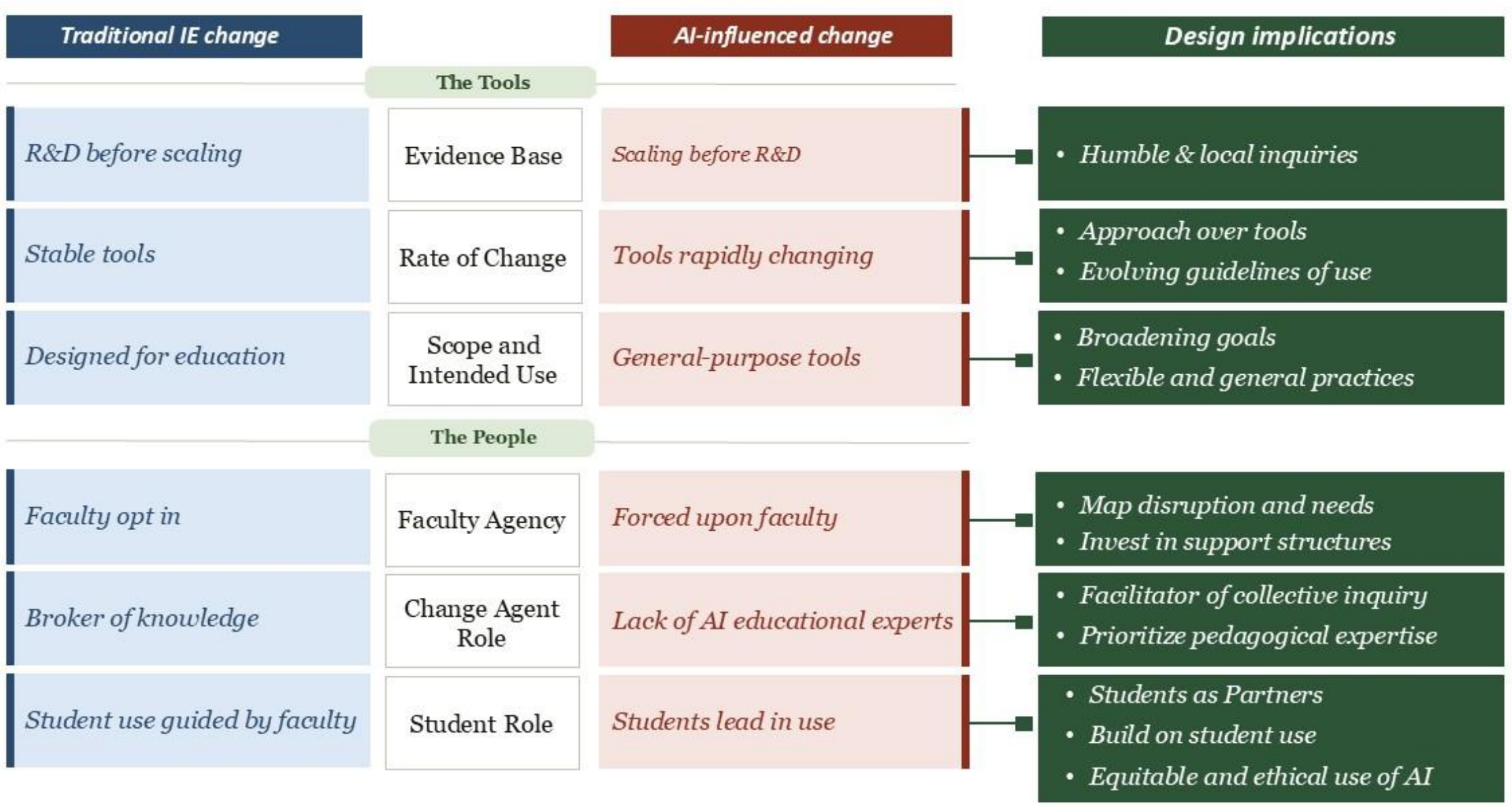

*Figure 2: The full framework. For each of the six dimensions, the figure summarizes (a) the assumption embedded in prior IE change models, (b) how that assumption is disrupted by generative AI, and (c) the design implications we derive for AI-driven institutional change.*

## 4. The Framework in Practice

We illustrate the utility of the framework through a brief case-study of a faculty workshop series we designed and facilitated in our CU physics department in Spring 2026. The department, housed in a large public R1 research university, has a strong tradition of systematic education reform – having participated in many initiatives, including the SEI and DAT models described above. Faculty discussions about AI, initiated by the second author in the prior term, generated enough interest and motivation to turn into a structured workshop series. It comprised six bi-weekly, 60-min sessions, with approximately ten faculty attending any given session and nineteen participating across the series. Our goals were to respond to the arrival of AI *as a department* - supporting both faculty and students - and to begin establishing shared departmental resources and policies.

Guided by our framework, we framed the series as collective, local inquiry rather than general faculty training on AI tools: each session presented locally sourced data or materials, had faculty test and reflect on them in small groups, and closed with whole-group discussion. The six sessions addressed course policies around AI; discussing AI use with students; homework corrections with and without AI; using AI in course tasks (generating code, diagrams, simulations, and explanations); AI and assessment; and a closing synthesis. Materials collected and developed across sessions were assembled into a departmental repository for faculty use, to be iterated and expanded as our knowledge develops. Our aim here is not to report workshop outcomes or validate the framework, but to show how its dimensions and implications shaped concrete design decisions. Table 1 pairs each dimension to a specific decision that shaped our workshop design, and to an example outcome. Workshop materials are available from the authors upon request.

| **Dimension** | **Design decision** | **Example Outcomes – resources, tensions** |
|---|---|---|
| **Evidence base** *humble, local inquiry* | Rather than relying on AI-literacy frameworks, we anchored each session in departmental-sourced materials, and organized small-group discussions testing them. | A shared departmental repository including local policies, learning activities using AI, survey and clicker questions to engage students in AI policy discussions. |
| **Rate of change** *approaches over tools* | We organized sessions around pedagogical approaches (policy, norm-setting, homework correction guidance), not specific platforms. | Faculty were often concerned with specific tool comparison and capabilities, especially of institutionally licensed tools. The "approaches over tools" lens reminded us to treat these questions as secondary and shift the discussion to the underlying pedagogical approaches. |
| **Scope** *general practices* | We built our activities around the general-purpose tools students and faculty already use (e.g. Gemini/ChatGPT). | Faculty-sourced sample activities that make use of general-purpose platforms and promote general AI practices such as generating code, diagrams, or simulations, then discerning and revising output. |
| **Faculty agency** *map disruption* | We conducted faculty interviews before the workshop, to understand where faculty stood regarding AI. | Faculty-course map, on which interviewed instructors ranked the level of disruption AI introduced to their course goals and structures. Computational and academic writing courses were reported as highly affected, while advanced and graduate courses were seen as less so, marking the disruption heterogeneity. |
| **Change-agent** *facilitator* | We facilitated collective inquiry around curated resources rather than disseminating 'best practices'. | The shared session format – local data, small-group testing, reflective discussion – adapting DAT structures and creating departmental community structures. |
| **Student role** *build on use* | We led with student surveys and let findings drive session design. | A physics students survey of AI use and perceptions, which several faculty adopted. This generated data across five different courses (N≈350). |

Table 1: Applying the framework in practice. For each of the six dimensions, we provide an example of how it shaped a concrete decision when leading our faculty AI workshop series. The third column presents an outcome - a resource, data or tension - that emerged.

As a deeper dive into one of these dimensions, we share how student perspectives were integral; student survey data shaped the workshop sessions. In the first session we presented findings from a campus-wide survey (N≈3000). It showed that student AI use is widespread, but that students are also concerned about it and want guidance from their instructors on appropriate use in their courses. A leading reason students gave for not using AI to produce finished work was to respect their instructors' policies. These findings pointed to the importance of having AI policies and discussing AI use with students – which were the topics of our first two sessions, and practices largely absent among our faculty until then. In these meetings we shared existing local policies and materials to engage students in dialogue around AI use and perceptions.

One of the instruments we shared was a survey on AI use in coursework designed for physics students, along with results from one of our middle-division, majors' courses (N=37). Based on this discussion, several instructors were interested in issuing this survey in their courses, creating a dataset that we updated and shared across the sessions (reaching N≈350). One surprising result was that students reported "checking correctness" of their solutions as one of their most frequent uses of AI. In response, we designed our third session around this practice: we selected actual homework problems and student solutions from our courses. In small groups, faculty submitted these solutions to their preferred AI chatbot, tried different prompts and strategies for "checking correctness" of solution, and examined which approaches prompted reflection and which shortcut it. The discussion turned to whether and *how* we want students to use AI for checking their homework, and what guidance might redirect a widespread, unsupervised practice toward productive use.

Overall, the workshop case study reflects what reasoning with the framework might look like in practice: the dimensions helped us articulate and justify our design choices and supported departmental action under conditions of genuine uncertainty. For those interested, a forthcoming study will outline the workshop in detail and share the repository of materials collected.

## 5. Discussion and Conclusion

Prior models of institutional change designed to promote interactive engagement and student-centered practices are no longer sufficient to address the AI-infused educational landscape. As we have argued, the foundational difference lies in the distinction between AI as an *arrival technology* and prior IE reforms anchored on *adoption technologies* (Reich & Dukes, 2025) – a difference that challenges core assumptions embedded in our existing change frameworks.

Our framework identifies six dimensions along which those assumptions must be reconsidered: three concerning the tools at the center of change and three concerning the people who enact it. We do not claim these six dimensions are exhaustive; rather, they offer a starting point for developing change initiatives adapted to the conditions generative AI creates. Critically, building on prior models remains essential. As Reinholz et al. (2021) note, no single theory fits a particular change effort; different frameworks simultaneously illuminate different aspects of complex change processes. Our contribution is to point to central dimensions where AI-specific characteristics demand reconsideration, while preserving the accumulated insights of decades of IE institutional change research. Indeed, many of the design implications we derive – local inquiries, embedded change agents, student partnership – extend and adapt approaches already present in IE models (such as DAT, Reinholz et al., 2019, or DeLTA, Andrews et al., 2021). Our collective challenge is not to start over, but to adapt thoughtfully to the AI age. We propose our framework to work alongside prior change models, highlighting AI-specific considerations and pointing to critical adaption points.

It is worth noting that while we believe AI is influencing all courses, not all educational activities are necessarily directly impacted by it. Consider, for example, the canonical application of Peer Instruction (Mazur, 1997), where students engage in sense-making and argumentation in class without outside resources. In these low-stakes, formative assessments, students have little incentive and likely would not use AI tools (even if the opportunities exist). Other educational practices may fall into this 'AI-independent' category depending on how they are enacted, and we believe a systematic analysis of which activities are disrupted (or left undisturbed) by AI would be a valuable contribution to our field.

The disruption AI introduces is not uniform across the six dimensions. We conjecture that the tool-related dimensions are where IE and AI-driven change diverge most fundamentally, reflecting the disruption of AI as an *arrival technology*. It is in these dimensions that we should seriously rethink our change models, moving from dissemination of stable, evidence-based, and education-specific tools to approaches designed for genuine uncertainty. The people-related dimensions are more closely aligned: Many of the embedded expert core functions that proved valuable in IE reform (Chasteen et al., 2016; Huber & Hutchings, 2021) remain relevant for AI-driven change, even as their central role shifts from brokering known practices to navigating collective inquiry. Similarly, IE models that treat students as partners in reform (e.g., Otero et al., 2006; Wilson & Varma-Nelson, 2016) translate most directly into the AI context. Faculty agency shows greater divergence: while faculty in IE change could opt in, AI disruption to their classes is something they did not choose. However, many faculty have yet to recognize this disruption (nor are all courses equally disrupted). We can rely on IE-based strategies – e.g. facilitating faculty communication; partnering with faculty on course transformation (Chasteen et al., 2016) – to engage and sustain faculty in AI-driven transformation. The degree of alignment with or modification to traditional change strategies, of course, will depend upon the specific context and focus of change.

As becomes clear in the descriptions above, the dimensions we offered are not intended as orthogonal categories but are interrelated and interact in practice. For example, *scope and intended use* and *student role* are entangled, because AI's general-purpose nature is what drives students' adoption outside of institutional channels. *Faculty agency* is deeply tied to *change agent role* and actions. Nonetheless, each dimension points to distinct practical levers for change (i.e., the design implications), which is why we keep them analytically separate. Change efforts that consider multiple dimensions and their interconnectedness are likely to be more effective (Laursen & Austin, 2020; Reinholz & Apkarian, 2018).

The brief case study presented above illustrates how the framework can function in practice in the design of a faculty and staff workshop and highlights the key design considerations that follow our framework for AI-driven reforms (see Figure 2). Taken together, these design implications suggest a set of reorienting principles for AI-era institutional change: act with intellectual humility rather than premature confidence; invest in support structures for collective faculty inquiry rather than tool procurement; position change agents as facilitators rather than brokers of known best practices; build on students' existing AI practices and treat them as partners in the change process.

Several limitations of this framework should be acknowledged. First, our framework is theoretical: while we illustrate its application in a case study and draw from experiences with prior change efforts, we have not empirically tested whether change initiatives designed in accordance with the framework produce positive outcomes relative to alternative approaches. Such testing is an important direction for future research. Second, the framework is drawn primarily from U.S.-based change initiatives; the relative weight of the six dimensions, and specific design implications, may shift in different institutional contexts. Third, the AI landscape itself continues to evolve rapidly, and the dimensions most salient today may be reordered as the technology and our practices around it mature.

Although we have anchored our analysis in what we termed interactive engagement reform, the arrival of AI reshapes the landscape for STEM education reform broadly, including efforts concerning identity, equity, epistemology, and more. Working out how the framework dimensions apply to broader traditions - each with its own objects of change, communities, and relationship to AI - is a direction we hope others will pursue.

Lastly, we do not present this framework as exhaustive or final. Rather, we put it forward as a call for action, and as a starting point for individuals and groups to surface their assumptions and design choices. We can build on this initial framework, share approaches and engage in dialogue about how our change strategies evolve in light of AI impacts as we move forward with humility and intention into this uncertain landscape that has arrived.

**Declarations**

**Availability of data and materials**

The workshop materials described in this paper are available from the corresponding author.

**Ethics declaration**

Not applicable.

**Funding**

Not applicable.

**AI Disclosure**

Generative AI and machine learning tools were not used in the writing of the manuscript body. Claude (Anthropic) was used for copyediting and in the creation of Figures 1 and 2 with prompts to guide a draft of the figures' content, which was then edited by the authors.